# Intermolecular vibrational states far above the van der Waals minimum: combination bands of the polar $N_2O$ dimer

A.R.W. McKellar [a] and Moazzen-Ahmadi [b,*]

[a] Department of Physics and Astronomy, University of Calgary, 2500 University Drive North West, Calgary, Alberta T2N 1N4, Canada

[b] National Research Council of Canada, Ottawa, ON K1A 0R6, Canada

Address for correspondence:   Dr. N. Moazzen-Ahmadi
Department of Physics and Astronomy,
University of Calgary,
2500 University Drive North West,
Calgary, Alberta T2N 1N4,
Canada

* Corresponding author. Email: nmoazzen@ucalgary.ca (N. Moazzen-Ahmadi)

**Abstract**

Infrared combination bands of the polar isomer of the $N_2O$ dimer are observed for the first time, using a tunable infrared laser source to probe a pulsed slit-jet supersonic expansion in the $N_2O$ $v_1$ region ($\approx$2240 cm$^{-1}$). One band involves the torsional (out-of-plane) intermolecular mode and yields a torsional frequency of 19.83 cm$^{-1}$ if associated with the out-of-phase fundamental ($N_2O$ $v_1$) vibration of the $N_2O$ monomers in the dimer. The other band, which is highly perturbed, yields an intermolecular in-plane geared bend frequency of 22.74 cm$^{-1}$. The results are compared with high level *ab initio* calculations. The less likely alternate assignment to the in-phase fundamental would give torsional and geared bend frequencies of 17.25 and 20.16 cm$^{-1}$, respectively.

1. **Introduction**

Two isomers of the weakly bound nitrous oxide dimer have been experimentally observed by means of high-resolution infrared [1-10] and microwave [11] spectroscopy. The more stable low energy form is nonpolar with $C_{2h}$ symmetry, and has a centrosymmetric planar staggered parallel geometry. The higher energy form is polar with $C_s$ symmetry, having coplanar N$_2$O monomers pointing in the same direction, staggered, and approximately parallel. The energy difference between the zero point levels of the two isomers has been estimated to be 144 cm$^{-1}$ [12-14] or 137 cm$^{-1}$ [15] in high level *ab initio* calculations.

The intermolecular vibrations of N$_2$O dimer are of special interest for understanding intermolecular forces and for testing theoretical potential energy surfaces. Each isomer has four such fundamental modes, which can be described as: the van der Waals stretch, the out-of-plane bend or "torsion", and the two in-plane bends: contrarotary or "geared", and corotary or "antigeared". Three of these modes have been previously observed for the nonpolar isomer by means of infrared combination bands, but there has been no previous detection for the polar isomer. The experimental values for the nonpolar isomer are 27.3, 42.3, and 96.1 cm$^{-1}$ for the torsion, geared bend, and antigeared bend, respectively. These are in good agreement with recent *ab initio* theory [12,15]. The experimental values refer to the dimer in an excited (upper) state, for example, the N$_2$O $\nu_1$ stretch ($\approx$2220 cm$^{-1}$), because they are derived from infrared combination bands, whereas the theoretical values are for the intramolecular ground state. But there is good evidence that intermolecular modes of weakly-bound dimers are relatively insensitive to intramolecular excitation [16].

Intermolecular vibrations of weakly bound molecules like N$_2$O dimer tend to involve large amplitude motions, so harmonic estimates are not very useful and reliable theoretical predictions instead require detailed variational calculations in multiple dimensions using an accurate potential energy surface. But the intermolecular modes of the higher energy polar $(N_2O)_2$ isomer are embedded in an increasingly dense manifold of excited overtone and combination vibrations belonging to the lower energy nonpolar isomer. For this reason, identification and characterization of the polar isomer modes is theoretically challenging; many levels have to be calculated and then the desired polar isomer levels have to be identified and distinguished from the unwanted nonpolar ones. These difficulties have been addressed in two publications. Wang et al. [13] used wavefunction re-expansion (in terms of $J = 0$ eigenfunctions), and calculated transition intensities, to label the desired "bright" polar isomer states. Zheng et al. [15] instead used the nodal properties of the calculated wavefunctions for this purpose. For our purpose, the key results from these two papers are predicted values for the polar $N_2O$ dimer torsion and geared bend vibrations of 21.40 and 20.31 cm$^{-1}$ [13], or else 20.05 and 21.26 cm$^{-1}$ [15], respectively. These agree quite well with each other, except that the ordering of the two modes is reversed in the two calculations! Frequencies of the other intermolecular modes are expected to be significantly higher (stretch ≈ 44.17, antigeared bend ≈ 84.80 cm$^{-1}$) according to the predictions in [15].

In the present paper, we report detection of a weak torsional combination band of the polar isomer of $N_2O$ dimer in the in the $N_2O$ $\nu_1$ region. In addition to this band, fragments of another band are observed nearby, probably indicating a highly perturbed geared bend combination band. These results are of interest because they probe vibrational levels quite far (160 cm$^{-1}$) above the ground state of the dimer, in a region of increasing density of states (roughly one per cm$^{-1}$) where calculations become more difficult.

## 2. Results and analysis

Spectra were recorded using a previously described pulsed supersonic jet - tunable infrared quantum cascade laser apparatus [4,9,16]. The expansion gas was a mixture of about 0.16% $N_2O$ in helium, and the backing pressure was about 14 atm. Use of a dilute mixture is crucial for observing the higher energy polar $N_2O$ dimer. Reference gas ($N_2O$) and etalon spectra were recorded for frequency calibration, and the PGopher computer program [17] was used for spectral simulation and fitting, with the Mergeblends option used to fit blended lines to the intensity weighted average of the contributing transitions.

The observed spectrum is shown in Fig. 1. The portion shown in the lower panel was fairly easy to recognize as a torsional combination band, thanks to its similarity to the known torsional band of nonpolar $N_2O$ dimer (see Fig. 2 of [6] or [9]). A very good fit was obtained using the known polar $N_2O$ dimer ground state parameters [11] for the lower state, proving that it was indeed due to that isomer, and the observed *c*-type selection rules guaranteed that it was a torsional band. We assigned 93 observed lines in terms of 105 transitions (the difference is due to blended lines such as unresolved asymmetry doublets), with values of $J'$ and $K_a'$ ranging up to 11 and 4, respectively.

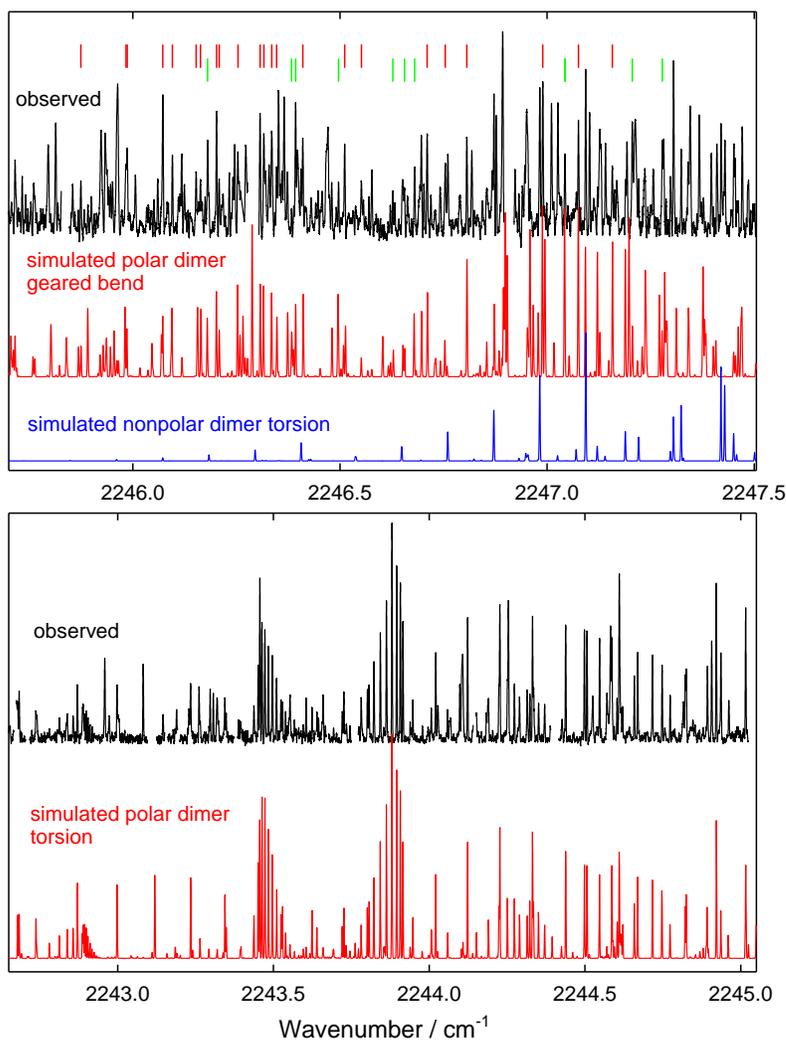

Fig. 1. Observed and simulated spectra of polar $N_2O$ dimer. Blanks in the observed traces correspond to regions of $N_2O$ monomer absorption. The lower panel shows the *c*-type torsional combination band, which is well fitted by the simulated spectrum. The upper panel shows the highly perturbed geared bend combination band, where only limited transitions with $K_a' = 0$ (red lines) and 1 (green) are assigned. The blue simulation shows part of the known nonpolar $N_2O$ dimer torsional combination band [6,9]. Note the different horizontal scales in the two panels. The observed absorption signal is magnified by a factor of 2.75 in the upper panel relative to the lower panel.

We first tried fitting the torsional band by itself, and the result was quite satisfactory, with an rms deviation of 0.0011 cm$^{-1}$. But this fit yielded a rather large change in the excited state $A'$ constant, with $A' \approx 0.2624$ cm$^{-1}$ compared to the ground state value $A'' = 0.3092$ cm$^{-1}$ [11]. As well, it gave a large negative value for the upper state distortion constant, $\Delta_K' \approx -0.0005$ cm$^{-1}$. The change in $A$ is a clear signal of Coriolis interaction, just as found previously for the analogous torsional combination band of nonpolar N$_2$O dimer, where it was ascribed to the $a$-type Coriolis with the higher lying geared bend mode [7,9]. The same explanation can be invoked here, and it basically tells us that the geared bend must lie above the torsion in the present polar N$_2$O dimer case.

The upper panel of Fig. 1 shows the spectral region just above that of the lower panel, where we might expect to observe this geared bend combination band. There is indeed a mass of weaker lines, but they are very irregular and attempts to assign them were unsuccessful at first. Eventually we discovered a number of line spacings which agreed with known polar dimer ground state combination differences, and were consistent with $a$- and $b$-type transitions as expected for the geared bend. The more convincing of these transitions involved upper state levels with $K_a' = 0$, while others involved the lower components of $K_a' = 1$, that is $(J, K_a, K_c) = (1,1,1), (2,1,2), (3,1,3)$, etc. But this was as far as we were able to go. The tentative assignments then involved 23 lines with $K_a' = 0$, $J' = 0$ to 7, and 11 lines with $K_a' = 1$, $J' = 1$ to 4.

Next, we tried a combined fit of the two bands which included the expected $a$- and $b$-type Coriolis interactions between them. These introduce matrix elements given by $\langle$tors, $k|H_a|$gb, $k\rangle = \xi_a \times k$, and $\langle$tors, $k|H_b|$gb, $k \pm 1\rangle = \frac{1}{2} \xi_b \times [J(J+1) - k(k \pm 1)]^{1/2}$, where tors and gb stand for the torsional and geared bend vibrations, $\xi_a$ and $\xi_b$ are the Coriolis parameters, and $k$ is the signed value of $K_a$. This fit worked very well for the torsional band: the rms deviation was improved from 0.0011 to 0.0003 cm$^{-1}$, the new $A'$ constant ($\approx$ 0.2913 cm$^{-1}$) was much closer to the ground state value, and the $\mathit{\Delta}_K'$ constant could simply be fixed at its ground state value. As well, the derived $a$- and $b$-type Coriolis parameters were quite reasonable. However, the new fit did not help to make any further sense of the supposed geared bend band, where the rms deviation was 0.0009 cm$^{-1}$ for the limited number of assigned lines. The parameters from this fit are listed in Table 1, and used for the simulated spectra in Fig. 1. Those for the geared bend band should be regarded with caution since they do not represent the band well, except perhaps for $K_a' = 0$.

The torsional band simulation in the lower panel of Fig. 1 works very well. The geared bend simulation in the upper panel looks vaguely reasonable in a qualitative sense, but as mentioned it only explains some transitions with $K_a' = 0$ (indicated with red lines) and a few with $K_a' = 1$ (green lines). We conclude that the geared bend combination band must be heavily perturbed by another vibrational state or states.

Table 1. Molecular parameters for polar $N_2O$ dimer (in $cm^{-1}$) [a]

|  | Ground State [11] | Torsion combination | Geared bend combination [b] |
|---|---|---|---|
| $\nu_0$ |  | 2243.7076(2) | 2246.6131(2) |
| $A$ | 0.3091713 | 0.29132(42) | 0.35616(69) |
| $B$ | 0.0541240 | 0.05280(35) | 0.05356(31) |
| $C$ | 0.0459192 | 0.046432(32) | 0.043589(55) |
| $10^5 \times \Delta_K$ | 1.267 | [1.267] [c] | [1.267] [c] |
| $10^6 \times \Delta_{JK}$ | -2.7359 | -7.3(33) | [-2.7359] [c] |
| $10^6 \times \Delta_J$ | 0.3037 | 1.14(14) | [0.3037] [c] |
| $10^6 \times \delta_K$ | 0.6338 | [0.6338] [c] | [0.6338] [c] |
| $10^7 \times \delta_J$ | 0.5945 | 5.1(10) | [0.5945] [c] |
| $\xi_a$ |  | 0.3000(24) |  |
| $\xi_b$ |  | 0.045(10) |  |

[a] Quantities in parentheses correspond to $1\sigma$ from the least-squares fit, in units of the last quoted digit.

[b] The parameters for this state are uncertain and not too meaningful because they only represent a limited number of assigned transitions.

[c] These centrifugal distortion parameters were fixed at ground state values.

## 3. Discussion and Conclusions

So far in this paper, an important fact has been ignored: there are *two* fundamental vibrations of polar $N_2O$ dimer in the $N_2O$ $\nu_1$ region. The stronger one at 2226.453 cm$^{-1}$ is due to in-phase vibration of the two $N_2O$ monomers, and was first to be observed [4]. The weaker fundamental at 2223.875 cm$^{-1}$ is the out-of-phase vibration [8]. So we actually expect two torsional and two geared bend combination bands. Based on the theoretical predictions [13,15] mentioned above, all four bands should lie in the 2243 – 2248 cm$^{-1}$ region. Instead, we observe only one well behaved torsional band, and one perturbed and ambiguous geared bend band.

The 2226.453 cm$^{-1}$ in-phase fundamental is observed to be about five times stronger than the 2223.875 cm$^{-1}$ out-of-phase one [8]. But combination band intensities depend on subtle effects of mechanical and electrical anharmonicity, so we should not necessarily associate the observed combination bands with the in-phase fundamental just because it is stronger. Indeed, for the nonpolar $N_2O$ dimer, the combination of the forbidden ($A_g$ symmetry) intramolecular fundamental and the intermolecular torsion ($A_u$) gives the allowed ($B_u$) torsional combination band [6], while the combination of the allowed ($B_u$) intramolecular stretch and the torsion gives a forbidden ($B_g$) combination band. The $C_{2h}$ symmetry of the nonpolar dimer allows these distinctions to be made, but for polar $N_2O$ dimer, which only has a plane of symmetry, there is no such help and all vibrational transitions are nominally allowed.

However, what if the $N_2O$ monomers in the polar dimer were side by side, and thus equivalent, rather than staggered? In this case, the dimer would have $C_{2v}$ symmetry, the in-phase fundamental and geared bend would be $A_1$, the out-of-phase fundamental would be $B_2$, and the torsion would be $A_2$. Then the torsional combination band associated with the in-phase fundamental would be $A_2$, which is forbidden, and the torsional combination band associated with the out-of-phase fundamental would be $B_1$, giving *c*-type selection rules. The geared bend combination bands associated with the in- and out-of-phase fundamentals would be $A_1$ (*b*-type) and $B_2$ (*a*-type), respectively. The analogy with the nonpolar dimer, and the behavior of the polar dimer in the $C_{2v}$ limit, both support the idea that the out-of-phase fundamental is more likely to give rise to a strong torsional combination band in the present case.

The high level theoretical predictions for the torsional frequencies are 21.40 [13] or 20.05 cm$^{-1}$ [15], while our observed torsional frequency is either 17.25 or 19.83 cm$^{-1}$, depending on whether it is associated with the in-phase or out-of-phase vibration. This would also seem to favor the out-of-phase assignment. The theoretical geared bend frequencies are 20.31 [13] or 21.26 cm$^{-1}$ [15], while our observed (but tentative) value is either 20.16 (in-phase) or 22.74 (out-of-phase) cm$^{-1}$. The situation is thus rather uncertain. However, we can be fairly sure that the observed torsional vibration lies below its geared bend counterpart, because of the observed $A'$ value as explained above, confirming the ordering predicted by Zheng et al. [15]. Also, we think that the two observed combination bands are probably associated with the same fundamental since they seem to be strongly linked by the Coriolis interactions.

It is easy to understand why the geared bend vibration might be highly perturbed. Assume equal or similar intermolecular frequencies for both the in- and out-of-phase cases, and associate the observed torsional band with the out-of-phase fundamental. Then the in-phase torsional combination vibration would fall at 2246.29 cm$^{-1}$, almost on top of the observed (perturbed) geared bend at 2246.61 cm$^{-1}$. Relatively modest Coriolis interaction parameters connecting the in-phase torsion and the out-of-phase geared bend could then result in large and irregular perturbations, as observed (the mixing parameters might not be too large because "in-phase" and "out-of-phase" represent separate manifolds, in some sense). We have tried to model this and other possible perturbation scenarios, but made no progress, due to the congestion and limited signal-to-noise ratio of the spectrum in the region of the top panel of Fig. 1.

In conclusion, intermolecular vibrations of the polar $N_2O$ dimer have been probed by observing two combination bands in the region of the $N_2O$ $\nu_1$ stretch fundamental. Since the polar isomer of the dimer lies about 140 cm$^{-1}$ above the more stable nonpolar isomer, these polar intermolecular vibrations occur in a range containing many nonpolar background vibrations, and this poses challenges for accurate theoretical predictions, even for the ground state where intramolecular modes are not excited [13,15]. The present observation, with the $N_2O$ $\nu_1$ intermolecular mode excited, effectively doubles the density of states since there are two such vibrations for the dimer (in-phase, out-of-phase). If associated with the out-of-phase fundamental, as we prefer, the clear and unambiguous *c*-type band at 2243.708 cm$^{-1}$ yields a torsional frequency of 19.83 cm$^{-1}$ for the polar dimer, in rather good agreement with previous theoretical predictions of 21.40 [13] or 20.05 cm$^{-1}$ [15]. Alternately, if associated with the in-phase fundamental, the observed torsional frequency is 17.25 cm$^{-1}$. The more ambiguous and only partially analyzed *a*- and *b*-type geared bend band at 2246.613 cm$^{-1}$ is most likely associated with the same fundamental as the torsional band, given the relatively strong Coriolis interaction between them. This gives an experimental geared bend frequency of 22.74 cm$^{-1}$, as compared to predictions of 20.31 [13] or 21.26 cm$^{-1}$ [15]. The alternate in-phase assignment gives 20.16 cm$^{-1}$ for the geared bend. In either case, the observed spectrum supports the ordering predicted by Zheng et al. [15], with the geared bend lying above the torsion.

**Acknowledgments**

We gratefully acknowledge the financial support of the Natural Sciences and Engineering Research Council of Canada. We thank A.J. Barclay for assistance with the experiment.

**References**

[1] Z.S. Huang, R.E. Miller, J. Chem. Phys. **89** (1988) 5408.

[2] H.-B. Qian, W.A. Herrebout, B.J. Howard, Mol. Phys. **91** (1997) 689.

[3] Y. Ohshima, Y. Matsumoto, M. Takami, K. Kuchitsu, Chem. Phys. Lett. **152** (1988) 294.

[4] M. Dehghani, M. Afshari, Z. Abusara, N. Moazzen-Ahmadi, A.R.W. McKellar, J. Chem. Phys. **126** (2007) 164310.

[5] M. Dehghany, M. Afshari, R.I Thompson, N. Moazzen-Ahmadi, A.R.W. McKellar, J. Mol. Spectrosc. **252** (2008) 1.

[6] M. Dehghany, M. Afshari, N. Moazzen-Ahmadi, A.R.W. McKellar, Phys. Chem. Chem. Phys. **10** (2008) 1658.

[7] M. Dehghany, M. Afshari, Z. Abusara, C. Van Eck, and N. Moazzen-Ahmadi, J. Mol. Spectrosc. **247** (2008) 123.

[8] G.M. Berner, A.L.L. East, M. Afshari, M. Dehghany, N. Moazzen-Ahmadi, A.R.W. McKellar, J. Chem. Phys. **130** (2009) 164305.

[9] M. Rezaei, K.H. Michaelian, N. Moazzen-Ahmadi, J. Chem. Phys. **136** (2012) 124308.

[10] R. Zheng, J.-B. Yang, D.-P. Yang, Spectrosc. Letters **48** (2015) 198-212. 1200 region

[11] N.R. Walker, A.J. Minei, S.E. Novick, A.C. Legon, J. Mol. Spectrosc. **251** (2008) 123.

[12] R. Dawes, X.-G. Wang, A.W. Jasper, T. Carrington Jr., J. Chem. Phys. 133 (2010) 134304.

[13] X.-G. Wang, T. Carrington, Jr., R. Dawes, A.W. Jasper, J. Mol. Spectrosc. **268** (2011) 53.

[14] J. Brown, X.-G. Wang, T. Carrington, Jr., Phys. Chem. Chem. Phys. **15** (2013) 19159.

[15] L. Zheng, Y. Lu, S.-Y. Lee, H. Fu, M. Yang, J. Chem. Phys. **134** (2011) 054311.

[16] N. Moazzen-Ahmadi, A.R.W. McKellar, Int. Rev. Phys. Chem. **32**, 611-650 (2013).

[17] PGOPHER, a Program for Simulating Rotational Structure, C.M. Western, University of Bristol, U.K., http://pgopher.chm.bris.ac.uk.